\documentclass[twocolumn,aps,prb,epsf]{revtex4}
\usepackage{amsmath}
\usepackage{epsfig}
\usepackage{epsf}
\usepackage{array}

\begin{document}

\title{Theory of Mott insulator/band insulator heterostructure}
\author{Satoshi Okamoto and Andrew J. Millis}
\affiliation{Department of Physics, Columbia University, 538 West 120th Street, New York,
NY 10027}
\date{\today}

\begin{abstract}
A theory of heterostructures comprised of LaTiO$_3$ (a Mott insulator) and
SrTiO$_3$ (a band insulator) is presented. The band structure of the Ti $d$%
-electrons is treated with a nearest neighbor tight-binding approximation;
the electric fields arising from the La$^{3+}$/Sr$^{2+}$ charge difference
and the carriers are treated within a Hartree approximation; and the on-site
interactions are treated by unrestricted Hartree-Fock. The phase diagram as
a function of interaction strength and layer number is determined and
predictions are made for optical conductivity experiments. A note worthy
finding is that the edges of the heterostructure are generally metallic.
\end{abstract}

\maketitle




\section{Introduction}

``Correlated electron systems'' (such as transition metal oxides) are
materials in which strong electron-electron or electron-lattice interactions
produces behavior incompatible with the standard ``density functional plus
Migdal-Eliashberg'' theory which describes most compounds. The past decade
has seen tremendous progress in the physics and materials science of
correlated-electron systems. Significant improvements in crystal and film
growth, in measurement techniques and in theory have led to a much improved
understanding of the bulk properties of these materials. An important
finding is that correlated electron systems exhibit a multiplicity of
interesting phases (superconducting, magnetic, charge and orbitally ordered)
along with associated novel excitations. For recent reviews, see Ref.~%
\onlinecite{Imada01}, or the articles in Ref.~\onlinecite{Tokura00}.

The recent success in treating bulk properties suggests that the time is
ripe for a systematic study of the surface and interface properties of
correlated electron systems. In addition to its basic importance as a
fundamental question in materials science, correlated electron
surface/interface science should provide the necessary scientific background
for study of potential devices exploiting correlated electron properties,
because essentially any device must be coupled to the rest of the world via
motion of electrons through an interface, and for study of correlated
electron nanostructures, because essentially the defining property of a
nanostructure is a high surface to volume ratio. The fundamental interest of
bulk correlated electron materials lies in the novel phases they exhibit,
and we therefore suggest that the fundamental issue for the nascent field of
``correlated electron surface science'' is `` how does the electronic phase
at the surface differ from that in the bulk''; in other words, ``what is the
electronic surface reconstruction.''

This question has begun to attract experimental attention. Hesper and
co-workers have shown that the [111] surface of K$_3$C$_{60}$ differs from
bulk because of charge transfer caused by a polar surface.\cite{Hesper00}
Matzdorf and collaborators have demonstrated that in the correlated electron
system Ca$_{0.9}$Sr$_{0.1}$RuO$_3$ (which exhibits Mott metal-insulator
transition), the surface layers remain metallic down to a lower temperature
than does the bulk system.\cite{Matzdorf00} Izumi and co-workers have
fabricated ``digital heterostructures'' composed of different transition
metal oxides and have demonstrated changes in electronic phase and other
properties depending on the thicknesses of different layers.\cite{Izumi99}
In an experimental tour-de-force, Ohtomo, Muller, Grazul, and Hwang have
demonstrated the fabrication of atomically precise digital heterostructures
involving a controllable number $n$ of planes of LaTiO$_3$ (a
correlated-electron Mott-insulating material) separated by a controllable
number $m$ of planes of SrTiO$_3$ (a more conventional band-insulating
material) and have measured both the variation of electron density
transverse to the planes and the dc transport properties of the
heterostructure.\cite{Ohtomo02} Their work opens the door to controlled
studies both of correlated electron physics in confined dimensions and of
the behavior of the interface between a correlated system and a uncorrelated
one.

Many physics and material science issues arise in considering the behavior
of correlated electrons near surfaces and interfaces. Atomic reconstruction
may occur, and may change the underlying electronic structure. For example,
the authors of Ref.~\onlinecite{Matzdorf00} argue that a change in tilt
angle of the surface RuO$_6$ octahedra increases the electronic hopping,
thereby allowing the metallic phase to persist to lower $T$. Also, as noted
e.g. by Hesper and collaborators, a change in structure will lead to changes
in Madelung potentials, and to the screening which helps define the values
of many-body interaction parameters.\cite{Hesper00} ``Leakage'' of charge
across an interface may change densities away from the commensurate values
required for insulating behavior. Substrate induced strain is well known to
change the behavior of films.\cite{Izumi01}

Sorting out the different contributions and assessing their effect on the
many-body physics is a formidable task, which will require a sustained
experimental and theoretical effort. The experiment of Ohtomo \textit{et al.}
offers an attractive starting point. In this system, the near lattice match
(1.5~\% difference in lattice parameter) and chemical similarity of the two
components (LaTiO$_3$ and SrTiO$_3$) suggests that atomic reconstructions,
strain, and renormalizations of many-body parameters are of lesser
importance, so the physical effects of electronic reconstruction can be
isolated and studied. Further, the near Fermi surface states are derived
mainly from the Ti $d$-orbitals,\cite{Saitoh95} and correspond to narrow
bands well described by tight-binding theory. However, the orbital
degeneracy characteristic of Ti provides an interesting set of possible
ordered phases.

In this paper we undertake a theoretical analysis of the correlated electron
behavior to be expected in lattice-matched digital heterostructures of the
type created by Ohtomo \textit{et al}.\cite{Ohtomo02} We focus on electrons
in the Ti-derived bands, and include the effects of the long-ranged electric
fields arising both from the La atoms and the electronic charge
distribution. We treat the extra on-site interactions via a Hartree-Fock
interaction. We calculate the electronic phase diagram as a function of
on-site interaction parameter and number of La layers and for the relevant
phases determine the spatial variation of charge, spin and orbital
densities. We obtain a complex set of behaviors depending on interaction
strength and number of La layers. Generally, we find a crossover length of
approximately three unit cells, so that units of six or more LaTiO$_3$
layers have a central region which exhibits bulk-like behavior. The outmost $%
\sim3$ layers on each side are however metallic (in contrast to the
insulating behavior of bulk LaTiO$_3$). For very thin superlattices the
ordering patterns differ from bulk. We calculate optical conductivity
spectra and show that this is a revealing probe of the electronic structure.

The rest of this paper is organized as follows: Section II defines the
model, parameter values, and method of study. Section III presents our
results for the $T=0$ phase diagram as a function of interaction strength
and number of LaTiO$_3$ layers. Section IV presents our results for the
spatial dependence of the charge density in the small $U$ regime where
neither spin nor orbital ordering occur and provides, for this simpler case,
an overview of the general features of the electronic structure. Section V
studies the onset of spin and orbital order as the interaction and layer
thickness are increased. Section VI discusses in more detail the special
case 1 layer. Section VII study the interrelation between the metallic behavior 
and the subband structure, and a stability under parameter variation. 
Section VIII presents representative results for the optical
conductivity, and shows how these may be used to elucidate the electronic
structure. Finally, section IX presents a summary of our findings and
important future directions, and implications for the general questions of
correlated electron surface and interface science, and is designed to be
useful to readers uninterested in the details presented in sections II-VIII.

\section{Formalism}

Both LaTiO$_3$ and SrTiO$_3$ crystallize in the simple ABO$_3$ perovskite
structure \cite{Maclean79} (more precisely, very small distortions occur
which we neglect here) and as noted by Ref.~\onlinecite{Ohtomo02} the
lattice constants of the two materials are almost identical; $a_{LaTiO_3}
\simeq a_{SrTiO_3} = 3.9$~\AA , minimizing structural discontinuities at the
interface and presumably aiding in the growth of high quality digital
heterostructures.

\begin{figure}[tbp]
\epsfxsize=0.7\columnwidth \centerline{\epsffile{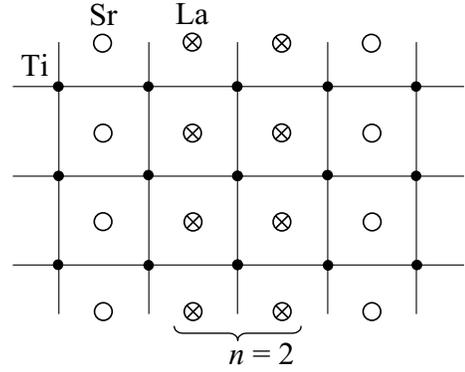}}
\caption{Schematic figure of the model used in the present study. Open and
crossed circles show the positions of Sr and La ions. (La layer number $n=2$%
) The dots show the positions of Ti ions. $x,y$ axes are chosen to be
parallel to the La plane, and $z$ axis is perpendicular to the plane.}
\label{fig:model}
\end{figure}

In this paper we consider an infinite crystal of SrTiO$_3$, into which $n$
adjacent [001] planes of LaTiO$_3$ have been inserted perpendicular to one
of the Ti-Ti bond directions, as shown in Fig.~\ref{fig:model}. We choose
the $z$ direction to be perpendicular to the LaTiO$_3$ planes, so the system
has a (discrete) translation symmetry in the $xy$ direction.

The relevant electronic orbitals are derived from the Ti $t_{2g}$-symmetry $d
$-states, and maybe labeled as $d_{xy}, d_{xz}, d_{yz}$. The derived bands%
\cite{Fujitani95} are to a good approximation given by a nearest neighbor
tight binding model with hopping parameter of magnitude $t \simeq 0.3$~eV
and spatial structure given by the Slater-Koster formula,\cite{Slater54} so
the $d_{xy}$ states disperse only in the $xy$ plane \textit{etc}.

We take the electrons to hop between Ti sites according to a nearest
neighbor hopping and to feel a potential defined by $(i)$ the Coulomb force
arising from the extra charge on the La relative to Sr $(ii)$ the Coulomb
force arising from the electrons on other Ti sites and $(iii)$ on-site
``Hubbard $U$ and $J$'' interactions with other electrons on the same site.
We take the form of the on-site interactions determined by Mizokawa \textit{%
et al}.\cite{Mizokawa95} and adopt values as discussed below. Thus 
\begin{eqnarray}
H_{tot}=\sum_a H_{hop}^{(a)} + \sum_i \bigl[H_{Coul}^{(i)} +
H_{on-site}^{(i)} \bigr]  \label{eq:Htot}
\end{eqnarray}
with 
\begin{eqnarray}
H_{hop}^{(xy)}= -2t \sum_{\vec k \sigma} (\cos k_x + \cos k_y) d_{xy, \vec k
\sigma}^\dag d_{xy, \vec k \sigma}  \label{eq:Hhop}
\end{eqnarray}
and similarly for $xz, yz$. We have 
\begin{eqnarray}
H_{Coul}^{(i)} = V_C(\vec R_i) \rho_d(\vec R_i)  \label{eq:Hcoul}
\end{eqnarray}
with $\rho_d(\vec R_i) = \sum_{a, \sigma} d_{a i \sigma}^\dag d_{a i \sigma}$
and 
\begin{eqnarray}
%
V_C(\vec R_i) = - \!\! \sum_{{\scriptstyle La \, sites} \atop {j}} 
\frac{e^2}{\varepsilon |\vec R_j^{La} - \vec R_i|}
+ \frac{1}{2} \sum_{{\scriptstyle Ti \, sites} \atop {j \ne i}}
\frac{e^2 \rho_d (\vec R_j)}{\varepsilon |\vec R_j - \vec R_i|}.
\label{eq:Vc}
\end{eqnarray}
Here the $\vec R_i$ are the positions of the Ti$^{(B)}$ sites in the ABO$_3$
lattice and $\vec R_j^{La}$ label the actual positions of the La ions, which
reside on the A sites.

We denote the dielectric function of the host lattice by $\varepsilon$. An
interesting issue arises here: SrTiO$_3$ is a nearly ferroelectric material.%
\cite{Sakudo71} The static dielectric constant becomes very large at long
wavelength and low temperatures, but $\varepsilon$ is much smaller at high
frequencies, room temperature, or short length scales. Also the polarization 
$P$ will increase more slowly at higher fields, and relevant quantity is $P/E
$. In this paper we have chosen $\varepsilon = 15$ as a compromise between
these effects. We discuss the consequences of different choices of $%
\varepsilon$ in the conclusion. We emphasize that incorporating the
ferroelectric tendencies of SrTiO$_3$ (including the associated lattice
distortions) in a more realistic manner is a important question for future
research.

Finally, the onsite $H$ is 
\begin{eqnarray}
H_{on-site}^{(i)}&=&U \sum_a n_{a i \uparrow} n_{a i \downarrow}
+(U^{\prime}-J) \sum_{a > b, \sigma} n_{a i \sigma} n_{b i \sigma}  \notag \\
&&+U^{\prime}\sum_{a \ne b} n_{a i \uparrow} n_{b i \downarrow} +J \sum_{a
\ne b} d_{a i \uparrow}^\dag d_{b i \uparrow} d_{b i \downarrow}^\dag d_{a i
\downarrow}  \notag \\
&&+J^{\prime}\sum_{a \ne b} d_{a i \uparrow}^\dag d_{b i \uparrow} d_{a i
\downarrow}^\dag d_{b i \downarrow} .
\end{eqnarray}
The critical issue is the strength of the on-site repulsion. For
definiteness we follow other studies which employ the ratios $U^{\prime}=
7U/9$ and $J=U/9$ which are very close to those determined by Mizokawa.\cite%
{Mizokawa95} Many workers have used the value $U \sim 5\mbox{-}6$~eV~$\sim
18t \mbox{-}20t$ estimated from high energy spectroscopies.\cite{Saitoh95}
However, optical conductivity studies of LaTiO$_3$ and related compounds
such as YTiO$_3$ find rather small gaps, in the range 0.2-1~eV,\cite%
{Okimoto95} suggesting $U \sim 2.5$~eV~$\sim 8t$. In view of this
uncertainty we investigate a range $U$ from $\sim 6t$-$20t$.

To study the properties of $H_{tot}$, Eq.~(\ref{eq:Htot}), we employ the
Hartree-Fock approximation replacing terms such as $n_{a i \sigma} n_{b i
\sigma}$ by $n_{a i \sigma} \langle n_{b i \sigma} \rangle + \langle n_{a i
\sigma} \rangle n_{b i \sigma}$; orbitally off-diagonal expectation values $%
(\langle d_{ai\sigma}^\dag d_{bi\sigma} \rangle)$ of the type considered by
Mizokawa and Mochizuki are stable only in the presence of a GdFeO$_3$ type
distortion which we do not consider. While not exact, the approximation
reveals the correct trends and in particular reveals insulating, ordered
states in the parameter regimes where these exist. To implement the
Hartree-Fock approximation we assume an initial distribution of site, spin
and orbital occupancies, obtain the one electron potential by factorizing
the interaction terms as described below, compute the band structure, obtain
new densities, and iterate until convergence is obtained for some $U$%
-values. Many iterations ($\sim 10^3$) are required to obtain well-converged
solutions, essentially because of the delicate balance of particle
distributions needed to screen the long range part of the Coulomb
interaction.

There are two types of solutions to the one-electron equations: bound
states, which decay as $|z| \rightarrow \infty$, and continuum states, which
do not. As usual in heterostructure problems the bound states give rise to
sub-bands, some of which are partially occupied. The ground state is
obtained by filling the lowest sub-bands up to the appropriate chemical
potential (determined by charge neutrality); the interaction-related terms
in $H_{Coul}^{(i)}$ and $H_{on-site}^{(i)}$ 
are then recomputed and the procedure is repeated until self-consistency is
obtained. Charge neutrality requires that the total density of electrons in
the bound-state sub-bands equals the total density of La ions. However, the
interplay between electron-La attraction and electron-electron repulsion
leads (in almost all of the cases we have studied) to a very weak binding of
the highest-lying electron states; indeed for large $U$ the Fermi level of
the partially filled sub-bands is only infinitesimally below the bottom of
the continuum bands. Also in some regions of the phase diagram, there are
several locally stable solutions and it is necessary to compare energies, to
determine the ground state. Between 200 and 3000 iterations were required,
with the larger number needed either when the Fermi level adjoins the
continuum states or when several different states are very close in energy.

We now mention some specific details of our solution to the Hartree-Fock
equations. We seek solutions which are bound in the $z$ directions; i.e.
functions $\phi(i,xy)$ which decay as layer index $i \rightarrow \pm \infty$%
, and which depend on the transverse $(xy)$ coordinates as discussed below.

For each orbital $\alpha$, the functions $\varphi^\alpha (i;xy)$ obey a
single particle Sch\"odinger equation which we write as 
\begin{eqnarray}
t_z^\alpha a^2 D_{2z} [\varphi^\alpha (i;xy)] + H_{xy}^{(i)} \varphi^\alpha
(i;xy) = E \varphi^\alpha (i;xy).
\end{eqnarray}
Here $D_{2z} \varphi$ is the discretized second order derivative $%
\varphi_{i+1} -2 \varphi_i+\varphi_{i-1}$ and $t_z^\alpha = t = 0.3$~eV for $%
\alpha = xz, yz$ and $=0$ for $\alpha = xy$. The term $H_{xy}^{(i)}$
contains the dispersion in the plane of the layers, as well as terms arising
from the long-ranges Coulomb interaction [Eq.~(\ref{eq:Hcoul})] and terms
arising from the Hartree-Fock approximation. These terms depend on the
on-site interactions and on the charge, spin and orbital density in layer $i$%
. A general mean-field state would break translation invariance in the $xy$
plane, in a manner whose amplitude would depend on layer index $i$, thereby
mixing the motion in $x,y$ and $z$ directions, leading to a very numerically
involved three dimensional self-consistency problem. In order to keep the
computations within reasonable bounds and explore wide ranges of parameters,
we have in most calculations restricted attention to state which preserve
translational invariance in the $xy$ plane. For these states the $xy$ and $z$
motions decouple, and physical quantities may be computed from the solution
of a 1$d$ problem combined with an integral over a two-dimensional momentum.
We also performed some investigations of states with a two-sublattice $xy$
plane structure. In this case, after Fourier transformation on the in-plane
coordinate the in-plane Hamiltonian may be written 
\begin{eqnarray}
H_{xy}^{(i)} = \left( 
\begin{array}{cc}
V_i + \Delta^i & \varepsilon_p \\ 
\varepsilon_p & V_i - \Delta^i 
\end{array}
\right)
\end{eqnarray}
with $V_i$ the part of the potential which is independent of sublattice, $%
\Delta^i$ the part which alternates between sublattices and $\varepsilon_p$
the in-plane dispersion in the reduced Brillouin zone. The variation of the
mixing term $\Delta^i$ with layer requires a separate treatment, lengthening
the numerical analysis considerably.

\section{Phase Diagram}

In this section we present and discuss the calculated phase diagram. The
geometry of the heterostructure distinguishes between the $xy$ orbital and
the $\{xz, yz \}$ orbitals. As noted above, we assume translation invariance
in the $xy$ plane. The symmetries which may be broken are therefore spin
rotation and orbital rotation ($xz$ and $yz$) and inversion along with
translation invariance in the $xy$ plane. Also, although it does not have
precise meaning except in the $n \rightarrow \infty$ limit, it is physically
sensible to interpret results in terms of $z$-direction translation symmetry
breaking when the spin or orbital density oscillates with respect to the
total charge density.

\begin{figure}[tbp]
\epsfxsize=0.8\columnwidth \centerline{\epsffile{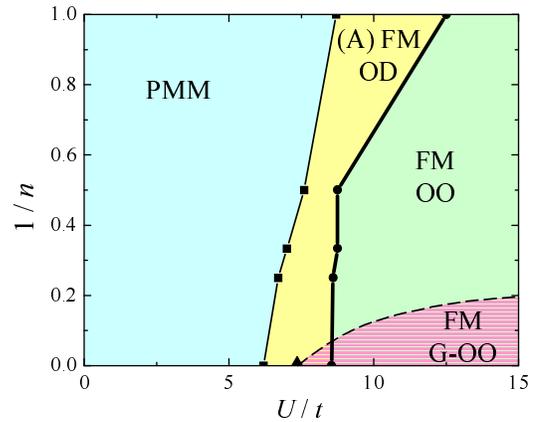}}
\caption{Theoretical ground state phase diagram as a function of the on-site
Coulomb interaction $U$ and inverse of the La layer number $n$. Parameter
values are chosen to be $U^{\prime}=7U/9$ and $J=U/9$. $e^2/(\protect%
\varepsilon a t) =0.8$ which corresponds to $\protect\varepsilon \sim 15$
with a lattice constant $a=3.9$~\AA and transfer intensity $t=0.3$~eV. The
triangle is the critical $U$ to the ($\protect\pi,\protect\pi,\protect\pi$)
orbital ordering at $n=\infty$ case. The broken line shows the expected
phase transition to the ($\protect\pi,\protect\pi,\protect\pi$)
antiferromagnetic orbital ordering at a finite $n$. }
\label{fig:diagram}
\end{figure}

Our calculated phase diagram is shown in Fig.~\ref{fig:diagram}. For reasons
of computational convenience in scanning a wide range of parameters, we
considered mainly phases with translation invariance in the $xy$ plane,
however for $n= \infty$ and $n=1$ we considered also an $xy$-plane two
sublattice symmetry breaking. We found in agreement with previous
calculations \cite{Mizokawa95,Ishihara02,Mochizuki02,Khaliullin02} that the
fully staggered phase is favored at $n=\infty$, but $xy$-plane symmetry
broken states could not stabilized in the one layer case. We have not yet
studied more general symmetry breakings for intermediate layer numbers, but
physical arguments presented below strongly suggests that these phases only
occur for larger numbers of layers ($n \agt 6$).

Four phases are found: a small $U$ phase with no broken symmetry, and
intermediate $U$ phase with in-plane translation-invariance spin order, but
no orbital order, and a large $U$ phase with both spin and orbital order.
The lower $U$ transition line varies smoothly with layer number, while the
larger $U$ transition is essentially independent of layer number for $n>1$.
Further, the $n=1$ intermediate $U$ phase is ferromagnetic whereas for $n>1$
the intermediate $U$ phase is antiferromagnetic. As will be shown in
sections below this behavior can be related to the sub-band structure but
the essential reason is that for small $n$ the charge density is spread in
the $z$ direction, so no layer has a density near 1.

\begin{figure}[tbp]
\epsfxsize=0.8\columnwidth \centerline{\epsffile{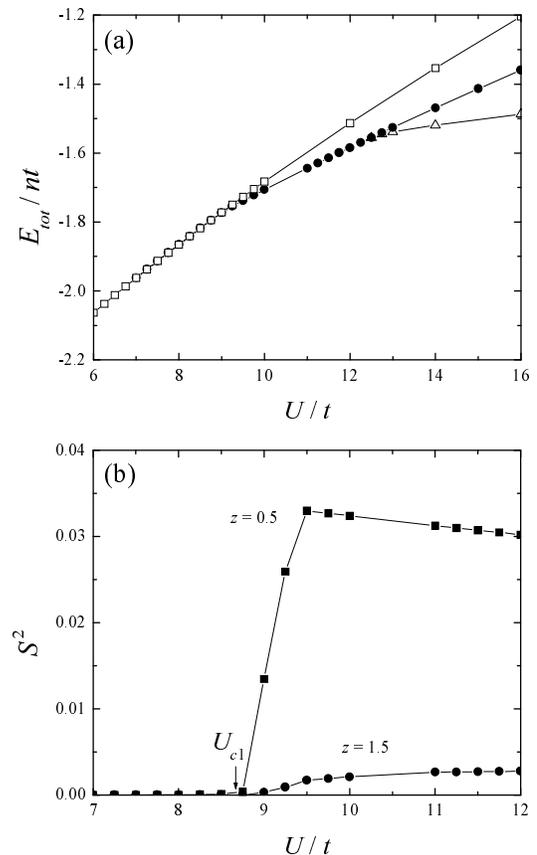}}
\caption{(a) Energy as a function of $U$ ($n=1$). Open squares: the spin and
orbital-disordered state, filled squares: the spin-order orbital-disorder
state, open circles: the spin and orbital-ordered state. (b) Square of
magnetization as a function of $U$ in the intermediate coupling region ($n=1$%
).}
\label{fig:energy}
\end{figure}

For all $n$, the lower-$U$ transition is found to be second order. As $U$ is
increased above this value, the magnetization increases rapidly and the
upper transition is strongly first order. Fig.~\ref{fig:energy} shows an
example of the procedure used to determine the location of the phase
boundary and the order of the transition. The upper panel of figure~\ref%
{fig:energy} displays the energies of several different phases. We identify $%
U_{c1}$ as the point where the different energies converge (this is most
easily done from a plot, not shown, of the energy differences on a expanded
scale) and $U_{c2}$ from the lowest crossing point of the energy curves.
That the energies of two qualitatively different states cross at $U_{c2}$
identifies this as a first order phase transition, that the energies merge
at $U_{c1}$ suggests a second order transition. Further evidence is provided
by panel (b), which shows, on an expanded scale, the magnetization as a
function of $U-U_{c1}$, and suggests a continuous decrease to zero.

The comparison to the $n=\infty$ limit is subtle. In bulk LaTiO$_3$ exhibits
a ($\pi,\pi,\pi$) type antiferromagnetic ordering. Theoretical calculations
(apparently confirmed by very recent NMR experiment, and x-ray and neutron
diffraction experiments)\cite{Kiyama03,Cwik03} suggest a four sublattice
structure which is very close to a ($0,0,\pi$)-type orbital ordering\cite%
{Mochizuki03} differing slightly from the ($0,0,\pi$) ordering studied here.
Stabilizing the observed state apparently requires a lattice distortion not
included in the model studied here. As $U$ is increased from zero the $n
\rightarrow \infty$ limit of the model considered here has a phase
transition which we believe to be of second order to an incommensurate
antiferromagnetic state with a wave vector which is an extremal spanning
vector of the Fermi surfaces of the bands arising from two of the orbitals
(say $xz,yz$) and which turns out to be very close to ($0,0,\pi$). (In fact,
for reasons of numerical simplicity we studied ($0,0,\pi$) ordering and
found a very weakly first order transition.) This transition is followed by
a strongly first order transition to one of a degenerate manifold of states
characterized by ferromagnetic spin order and ($\pi,\pi,\pi$) orbital order.
To maintain continuity with the heterostructure calculations we have also
suppressed the ($\pi,\pi,\pi$) ordering and located the phase boundary to
the (metastable) ($0,0,\pi$)-orbital spin-ferromagnetic state. We believe
the ($\pi,\pi,\pi$)-orbital spin-ferro state we have found is a reasonable
surrogate for the actual Mochizuki-Imada state found in experiment. Although
this state is favored at $n \rightarrow \infty$, the physical origin,
explained below in more detail, of our inability to stabilize states with
broken in-plane translation invariance at $n=1$ suggests that the ($%
\pi,\pi,\pi$) phases only occur for rather thick superlattices. The
essential point is that, for $n<6$, the solution in the large $U$ limit
consists of several partially filled sub-bands, which have effectively
minimized their interaction energy but which gain considerable kinetic
energy from motion in the $xy$ plane. Breaking of $xy$ translation symmetry
would reduce this kinetic energy gain without much decreasing the already
saturated interaction energy while $z$-direction kinetic energy is quenched
by the confining potential.

The estimate of $U = 8t {\mbox -} 10 t$ derived from optical measurements on
bulk, and from the recent work on the nature of the ordered phase suggests
an interesting series of phase transitions may occur as layer thickness is
varied.

\section{Small $U$ region}

In this section we present and discuss the results obtained for $U<U_{c1}$
where no symmetry breaking is observed. Although the system geometry implies 
$xy$ and $\{ xz,yz \}$ orbitals are inequivalent, the occupancies of those
orbitals are found to be almost the same; the ratio at the center cite is
given by $(n_{xz} - n_{xy})/n_{xz} = 0.023, 0.098$ and 0.049 for $n=1,2$ and
3, respectively, at $U = 6t$. Therefore, we mainly focus on the spatial
distribution of the total electron density below.

\begin{figure}[tbp]
\epsfxsize=0.8\columnwidth \centerline{\epsffile{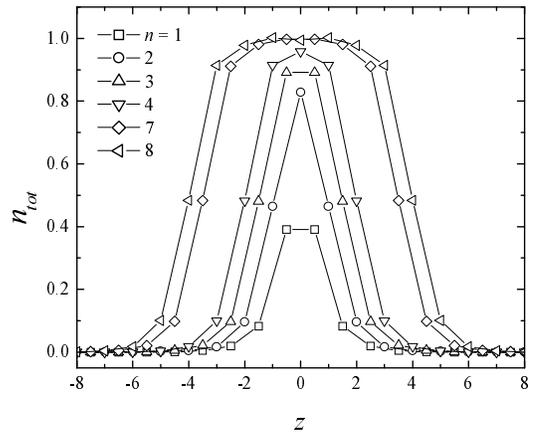}}
\caption{Calculated total electron density in the small $U$ region ($U=6t$).
Our results (not shown) for larger $U$ values are almost identical.}
\label{fig:n_tot}
\end{figure}

Numerical results for the electron density for several choices of La layer $n
$ and $U=6t$ are shown in Fig.~\ref{fig:n_tot}. Here we set the origin of $z$
axis at the center of $n$ La layers. Electron density is symmetrically
distributed around $z=0$. The density at the center site rapidly increases
with increasing $n$, and exceeds 0.9 at $n=6$. However, for this and larger $%
U$ values, the charge density never exceeds unity, even at the center sites.
Fig.~\ref{fig:n_tot} shows clearly that for thick layers the length scale
over which density vary is about three unit cells, and this number is almost
independent of $n$, and depends somewhat on $\varepsilon$ and weakly on $U$.%
\cite{U_dependence} Thus, at least 6 layers seem to be required before bulk
behavior is recovered in the central region. As is well known, SrTiO$_3$ is
close to ferroelectric, and the long wave-length, and linear response
dielectric constant becomes about 20000 at the lowest temperature.\cite%
{Sakudo71} The present results involve $\varepsilon$ at short scales, and
are clearly outside the linear response regime. However, temperature
dependent changes in the density profile should occur and are worth
experimental study. Another interesting feature to note is that for all $%
\varepsilon$ and $U$ studied the electron and hole distributions at large $n$
are almost symmetric at the edge $z=n/2$; $n_{tot}(z)$ and $1- n_{tot}(n/2-z)
$ fall on almost the same line (not shown).

It might be interesting to compare the present numerical results ($n=1,2$)
with the experimentally observed Ti$^{3+}$ spatial distributions.\cite%
{Ohtomo02} The observed Ti$^{3+}$ decay length of Ti$^{3+}$ is about 2~nm (5
unit cells) in both $n=1$ and 2, which is longer than the present
theoretical results (about 3 unit cells). The reported La-ion distribution
suggests that the spatial resolution of the EELS measurement is about 0.4~nm
(1 unit cell). Inclusion of such effect makes the density distribution
profiles closer to the experimental ones, but the experimental distributions
are still broader than the present theoretical ones. One possibility is
La/Sr interdiffusion (not considered here). An alternate origin could be a
chemical shift of the Ti $t_{2g}$ level. Here we have considered only the
long-range Coulomb energy, but an additional contribution arising from the
slightly different local structure [small ionic radius of La$^{3+}$ (1.045 
\AA ) than that of Sr$^{4+}$ (1.13 \AA )]\cite{Shanon76} could bring the
La-Ti $d$-levels closer to Sr-Ti $d$-levels. This effect has not been
included in this study.

\section{Onset of Polarization; Larger $U$}

This section presents results for large $U$ values, where spin and orbital
orderings are observed. The main focus is on how the orderings are developed.

\begin{figure}[tbp]
\epsfxsize=1\columnwidth \centerline{\epsffile{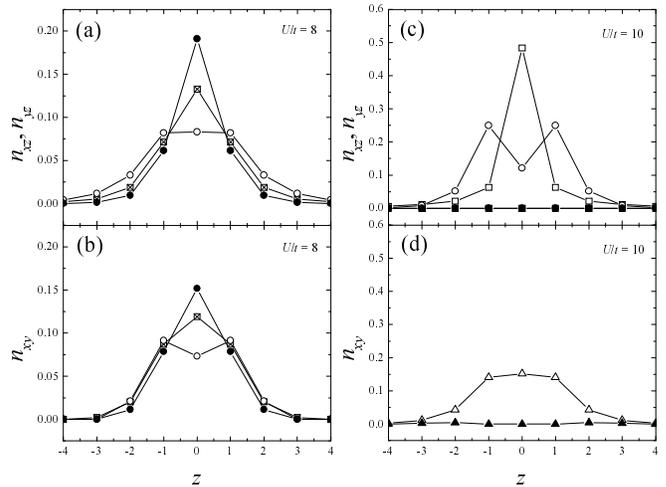}}
\caption{Electron density per spin and orbital in the intermediate coupling
regime $U=8t$ (a,b) and the strong coupling regime $U=10t$ (c,d) for a
heterostructure with two La layers ($n=2$). Open and filled symbols show the
densities of the up and down spin electrons, respectively. Circles and
squares in (a,c) are for the $xz$ and $yz$, respectively. Crossed squares in
(a,b) show the electron density per spin and orbital in a metastable
paramagnetic state. }
\label{fig:so_order}
\end{figure}

Figure~\ref{fig:so_order}~(a) and (b) show the electron densities per
orbital in the intermediate coupling regime ($U=8t$) with $n=2$ where only
spin ordering is observed. The difference between the $\{ xz,yz \}$ and $xy$
occupancy arise only from the symmetry. As a comparison, electron densities
per orbital and spin in the metastable paramagnetic state are also shown
(crossed squares). From the difference between the up and down electron
occupancy, one can notice that the spin density oscillates around the center
site. (although no site is fully polarized) In contrast for $n=1$ and $%
8t<U<12t$, the magnetic state is different: the spins align
ferromagnetically. The behavior at $n=1$ can be attributed to the lower
charge density at each site implying low occupancy of each sub-band, so a
small shift between the up and down electron sub-bands can stabilize the
fully polarized state. On the contrary, ferrimagnetic spin ordering at $n
\ge 2$ comes from a delicate balance between the confinement potential and
on-site Coulomb repulsion.

The electron densities per spin and orbital in the strong coupling regime ($%
U=10t$) are shown in Fig.~\ref{fig:so_order}~(c) and (d). The difference
between $xz$ and $yz$ orbital occupancy clearly indicates antiferro-type
orbital ordering along the $z$ direction ($0,0,\pi$). One can also notice
that spin polarization is almost complete. A tiny amount (less than 1 \%) of
down spin electrons exist only in the $xy$ orbitals. Note that the orbital
ordering in the strong coupling regime is different from that expected in
the bulk limit where translational symmetry in $xy$ plane would also be
broken. In the small $n$ region, the symmetry in $xy$ is expected to be
conserved because this symmetry is favorable to gain the kinetic energy by
two dimensional electron motion.

\section{$n=1$, possibility of staggered in-plane ordering}

In this section, we discuss special features of the $n=1$ case. Within the
Hartree-Fock approximation we have not found any stable states which break $%
xy$ plane translational invariance. We see from Fig.~\ref{fig:n_tot} that at
essentially all $U$ the occupancy of any particular orbital state is low.
For $U \agt 13t$ the orbital ordering in the $z$ direction (say $xy$
occupancy to the left of the La layer and $yz$ occupancy to the right),
along with the ferromagnetism effectively eliminates the on-site interaction
contribution to the Hartree-Fock energy, while the fact that the one
electron is shared between the two bands means that at each sub-band there
is a large kinetic energy gain of order $-\frac{2t}{\pi} \sin p_F$ per
sub-band. We believe this physics explains why we at $n=1$ have been unable
to to stabilize a state with two-sublattice in-plane symmetry breaking. In
bulk at large $U$, the kinetic energy is quenched and the ($\pi,\pi,\pi$)
state is stabilized relative to the ($0,0,\pi$) state by a superexchange of
approximately $2t^2/(U^{\prime}-J) \sim 3t^2/U$ but in the $n=1$
superlattice the band energy gain is approximately $-\frac{2t}{\pi} \sin p_F$%
. Indeed this argument suggests that until the band filling becomes close to
1 (which only happens for $n \agt 6$) the kinetic energy gain in $xy$%
-invariant states is larger than the superexchage energy gained by ordering
so that for a range of small $n$ (perhaps $n < 6$) staggered in-plane
ordering may be neglected.

We do expect the $xy$-translation invariant states we have found to be
unstable to some form of weak incommensurate density wave ordering, because
the resulting sub-band structure has a one-dimensional character from the $xz
$ and $yz$ sub-bands. However, this ordering is expected to be weak because
the residual interactions, not already included in our Hartree-Fock
solution, are not strong. For example, in the $U>13t$ regime, the on-site
interactions are fully quenched by the ferromagnetic ordering so
incommensurate charge density wave is driven only by the $2p_F$ component of
the long ranged Coulomb interaction, which we estimate from the sub-band
charge distribution with half width $\xi$ to be $V_c(2p_F) \approx C~(2
e^2/\varepsilon a)$ with $C = \int_0^\infty du \exp[-u^2(\pi \xi/a)^2]/\sqrt{%
1+u^2} \approx 0.3 \mbox{-}0.4$. For the parameters used here $V_c
\chi_{2p_F} \sim \lambda \ln(E_F/T)$ with $\lambda \ll 0.1$, so the
instability occurs at a very low $T$. We note, however, that the precise
critical behavior of the second order small $U$ transition found in our
Hartree-Fock approximation is likely to be complicated by the additional
effects of $2p_F$ instabilities in the quasi one dimensional bands. We leave
this interesting issue for future research.

\section{Metallic edge and stability under parameter variation}

The experimental results of Ohtomo \textit{et al}.\cite{Ohtomo02} 
show that the interface between the two types of insulators supports a metallic 
state. From numerical results for thick layer (Fig.~\ref{fig:n_tot}), 
one finds a central region where $n \simeq 1$, 
in which one expects bulk-like behavior, and an edge region with $n \ll 1$, 
which one might expect to exhibit metallic behavior. 
In this section, we study the interrelation between the subband structure and 
metallic behavior. 
We show that the edge region is responsible for the metallic behavior, 
and study the dependence of the metallic behavior on parameter changes. 

\begin{figure}[tbp]
\epsfxsize=0.7\columnwidth \centerline{\epsffile{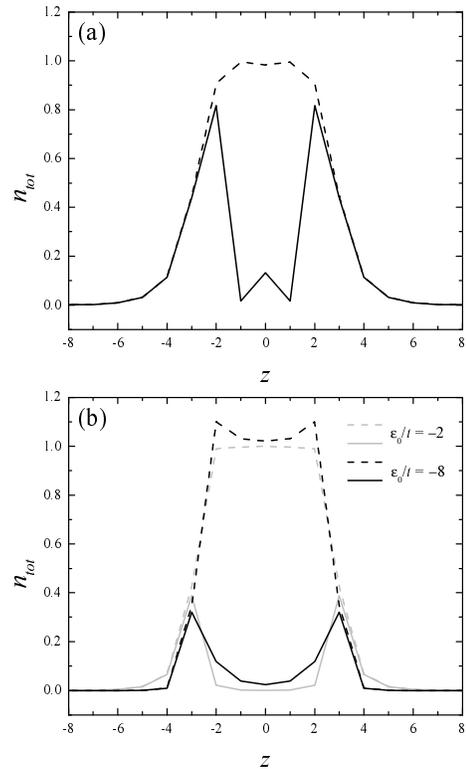}}
\caption{{\it Upper panel:} density profile (dashed line) and
density in partially occupied bands of heterostructure with 
$n=6$, $U/t=10$ and $\varepsilon = 15$ 
calculated in orbitally-ordered ferromagnetic phase. 
{\it Lower panel:} Density profile (dashed lines) and density in unoccupied bands
(light and heavy solid lines) for on-site binding to $Ti$ sites 
in heterostructure (on site energy for 
$Ti$) between two $La$ planes $\varepsilon_0$; on site energy for ``edge'' Ti
sites (between $Sr$ and $La$ planes) $\varepsilon_0/2$. }
\label{fig:nunoccu}
\end{figure}

Within the Hartree-Fock method we use here, 
physics of the metallic edge is manifested as 
follows. There are many bound-state solutions (wave functions decaying as 
$|z|$ is increased away form the heterostructure), 
whose dispersion in the in-plane direction gives rise to sub-bands. 
For thin heterostructures, all sub-bands are partially filled 
(implying metallic behavior in the heterostructure plane) 
whereas for thick heterostructures (in ordered phases), 
some sub-bands are fully filled and some are partly filled. 
The fully filled sub-bands have $z$-direction wave functions implying charge 
density concentrated in the middle of the heterostructure, whereas the 
partially filled bands have charge concentrated near the edges. 
The upper panel of Fig.~\ref{fig:nunoccu} shows the 
density profiles and the occupancy of partially filled bands for 
the heterostructure with $n=6$ and $U/t=10$ in a orbitally ordered 
ferromagnetic phase. 
We observe partially filled bands corresponding to metallic
behavior at the heterostructure edges where the density drops from 
$\sim 1$ to $\sim 0$. 

Two key assumptions underlying the calculations presented so far are 
the translation invariance of the on-site energy $\varepsilon _{0}$ 
(because in both components of the heterostructure the electronically active 
ion is octahedrally coordinated $Ti$) and relatively large value for the 
dielectric constant (to represent the nearly ferroelectric nature of 
$SrTiO_{3}$). 
We investigate the sensitivity of this result to changes in these parameters. 
The lower panel of Fig.~\ref{fig:nunoccu} shows the changes in the density 
profile and in the occupancy of partially filled bands occurring 
if the on-site energy is changed for sites near $La$ sites. 
Here, we have chosen the site energy of those $Ti$ sites which sit between
two $Sr$ planes to be zero, the energy of a $Ti$ site between two $La$
planes to be $\varepsilon _{0}<0$ and the energy of a $Ti$ site between a $La$ 
plane and a $Sr$ plane to be $\varepsilon _{0}/2$. 
For $\varepsilon_0 = 0$ several partially filled bands exist, 
leading to a high density of metallic electrons. As $\varepsilon_0$ is decreased, 
the electronic structure rapidly rearranges so that most bands become fully filled. 
However, even for the largest $\varepsilon_0$ the geometry ensures that one electron
always remains to be shared between two nearly degenerate bands, one on either side
of the heterostructure, so the metallic behavior remains robust. 

\begin{figure}[tbp]
\epsfxsize=0.7\columnwidth \centerline{\epsffile{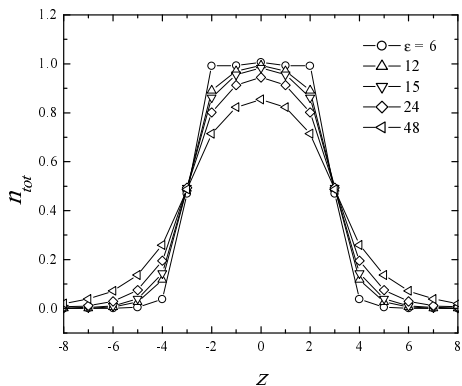}}
\caption{Dependence of total charge density on dielectric constant
calculated for paramagnetic phase of $6$ layer heterostructure 
with $U/t=10$ in paramagnetic phase.}
\label{fig:n6e_dep}
\end{figure}

We may similarly consider changes occurring as the dielectric constant is varied. 
Fig.~\ref{fig:n6e_dep} shows the paramagnetic phase density profile occurring as
the dielectric constant is varied over a wide range. The main effect is to 
decrease the tailing of charge density far into the $SrTiO_{3}$ layer. 
However, there again remains one electron to be shared between the two edge subbands, 
the basic phenomenon of partly filled bands at the edge is unaffected. 

Summarizing, for reasonable parameters we find that there are always partially 
filled bands, corresponding to metallic behavior at the heterostructure edges. 
The possibility of metallic 
behavior at the edge of a Mott insulator system appears to be confirmed by 
the experimental results of Ohtomo \textit{et al}.\cite{Ohtomo02} would be 
interesting study further both experimentally and by more sophisticated 
theoretical methods. 

\section{Optical Conductivity}

There are two classes of solutions to the Hartree-Fock equations: bound
states, which have wave functions which decay as $|z|$ is increased away
from the La layers, and continuum states, extended in all three directions.
Although the existence of sharply defined single-particle states at all
energies is an artifact of the Hartree-Fock approximation, we expect this
qualitative structure to survive in a more sophisticated treatment.
Therefore, in this section we use the Hartree-Fock approximation to show how
optical conductivity can provide information on the nature and filling of
the bound states. We also present a qualitative discussion of the effect of
interactions beyond Hartree-Fock.

We consider electric field directed perpendicular to the La-planes (parallel
to $z$) and use the Peierls-phase approximation to determine the current
operator 
\begin{eqnarray}
\hat J_z = -i t \sum_{i a=xz,yz} \bigl[ d_{a i \sigma}^\dag d_{a i+\hat z
\sigma} - h.c. \bigr].
\end{eqnarray}
We then evaluate the usual Kubo formula using the Hartree-Fock eigenstates.
Note that in the nearest neighbor tight binding approximation the $d_{xy}$
states do not couple to $z$-electric field, so we will not discuss them in
this section.

\begin{figure}[tbp]
\epsfxsize=0.8\columnwidth \centerline{\epsffile{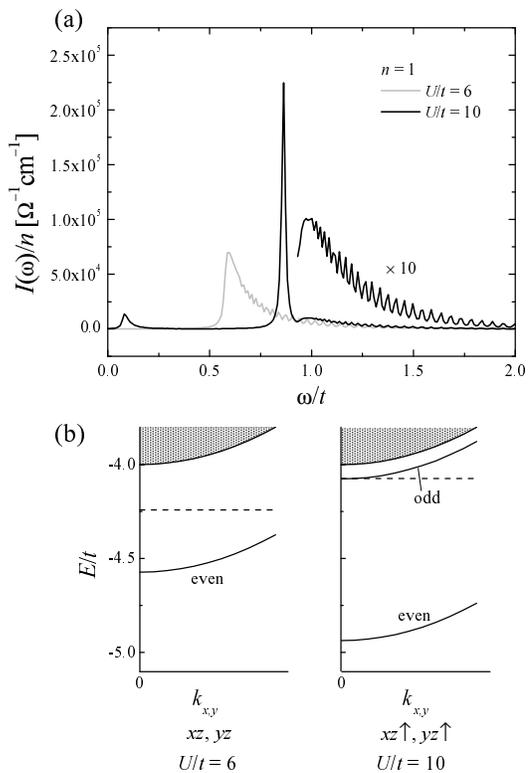}}
\caption{(a) Optical conductivity for $n=1$ with $U=6t$ (small $U$ regime)
and $10t$ (intermediate $U$ ferromagnetic state). Electric field $E$ is
chosen to be parallel to $z$. Delta functions are approximated as Lorentzian
with the half-maximum half-width $0.1t$. (b) Schematic illustrations of
sub-band energy structure at $U=6t$ (left) and $U=10t$ (right). Shaded areas
show the continuum state. Broken lines show the Fermi energy.}
\label{fig:spectra_n1}
\end{figure}

Figure.~\ref{fig:spectra_n1} shows the calculated $T=0$ conductivity for the
case of one La layer, and two values of $U$: $U=6t$ (light curve), for which
the system is in the spin and orbital disordered phase, and $U=10t$ (heavy
curve), for which the system is in the ferromagnetic, orbitally disordered
state.

For $U=6t$ the $xz,yz$ part of the spectrum (shown in the lower left panel)
consists of one four-fold degenerate bound-state sub-band, containing
approximately 1/6 electron per spin per orbital, as well as empty continuum
bands. The optical spectrum therefore consists of one feature, corresponding
to a bound-continuum state transition. The absorption is peaked at the bound
state energy, the width reflects the degree of overlap between continuum and
bound state energies. The oscillator strength is related to the kinetic
energy in the $z$ direction, and is approximately $10^4~\Omega^{-1}$cm$^{-1}$%
/La. We expect this spectrum to be very little affected by correlation
effects beyond Hartree-Fock, because the basic bound state energetics are
fixed by charge neutrality and Coulomb interaction, and the final states are
continuum states, which are delocalized in space.

Also shown in the upper panel of Fig.~\ref{fig:spectra_n1} is the spectrum
corresponding to $n=1$ and $U=10t$. For these parameters the electronic
structure has changed (lower right panel of Fig.~\ref{fig:spectra_n1});
there are now two sub-bands; one more strongly bound than in the $U=6t$ case
and holding more electrons (but still only approximately 1/3 filled) and one
only very slightly filled band. The calculated $\sigma(\omega)$
correspondingly exhibits three features, a weak low energy ($\omega \sim 0.1
t$) feature arising from transitions from the very slightly filled bound to
the continuum, a sharp higher energy ($\omega \sim 0.9 t$) feature arising
from the intersubband transition (allowed because the lower sub-band is even
under $z \leftrightarrow - z$ and the upper is odd) and a very broad feature
from lower bound-continuum transitions. The intersubband transition contains
about 2/3 of the total oscillator strength; the remainder is mainly in the
bound-state to continuum peak. We note that as expected on general grounds,
the presence of a bound-state to bound-state transition sharply reduces the
weight in the continuum (compare $U=8t,10t$), essentially because of the
requirement that the continuum states be orthogonal to both bound states.

Within the model we have used the intersubband transition is a delta
function, because lifetime effects have been neglected and the $x,y$ and $z$
direction dispersions decouple, so the 2 sub-bands disperse in the same way
as $k_x$ or $k_y$ is varied. The dispersions decouple because we have
adopted a nearest neighbor tight binding model, however this is believed to
be reasonably accurate in practice. A potentially more significant source of
broadening is the ``Hubbard $U$'' interaction. Experience from simpler
models suggests that for these $U$-values (of order of the critical value
for the bulk system Mott transition) and electron concentration (less than
0.4 electrons per layer, far from the Mott value $n=1$) these effects are
not too severe: essentially because the probability of 2 particle collisions
scales as $n^2$, only a small fraction of the spectral weight will be
shifted to an ``upper Hubbard band'' feature at an energy $\sim U$, and
lifetime broadening will be rather less than the Fermi energy, so roughly we
expect the peak to remain unchanged. These arguments apply also to the other
cases we consider below.

We now discuss the $n=2$ case. In the small $U$ paramagnetic, orbital
disordered case, the electronic spectrum exhibits two bound-state sub-bands;
arising from a more strongly bound even parity state and a weakly bound odd
parity state. The band structure is very similar to that shown in the right
panel of Fig.~\ref{fig:spectra_n1}~(b), but without spin polarization. The
optical conductivity (not shown) is essentially the same as the $U=10t$
curve shown in Fig.~\ref{fig:spectra_n1}. Increasing $U$ to $U=8t$ leads to
a spin ordered (here ferrimagnetic) orbitally disordered state, with four
sub-bands [three below $\mu$ and partially occupied] and correspondingly two
bound-state to bound-state transitions (one for each spin) and three
bound-state to continuum transitions (the middle one of which is evidently
extremely weak).

As a final example, we show the optical spectrum and energetics for the
spin-ferro orbital-antiferro $U=10t$ state. Five bound states now occur, and
correspondingly three sharp peaks (recall the optical selection rule allows
coupling only from even to odd parity states). Four bound-state-continuum
transition should be visible (the highest-lying $xz$ even parity state is
essentially unoccupied) but again all are weak and the lowest-lying (arising
from the odd $d_{xz}$ band) is especially weak because the low-lying
even-parity continuum states must be orthogonal to the even-parity bound
state, which absorbs all the oscillator strength.

\begin{figure}[tbp]
\epsfxsize=0.9\columnwidth \centerline{\epsffile{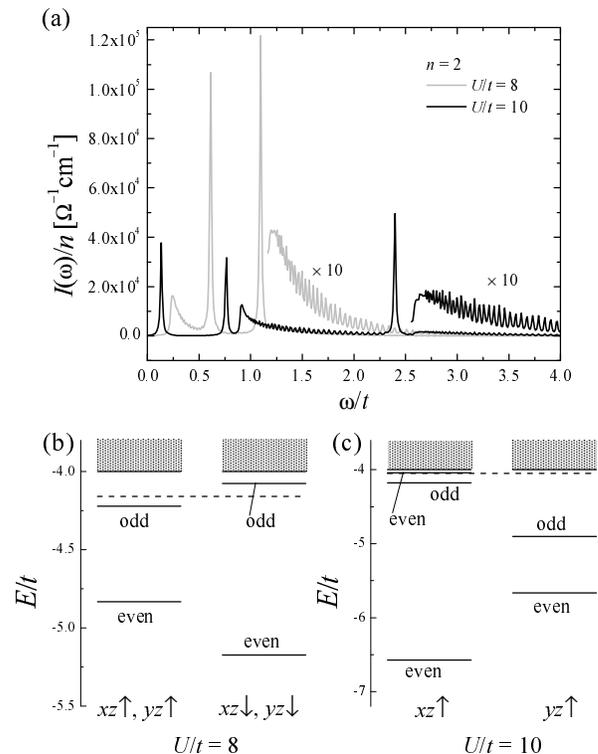}}
\caption{Optical conductivity for $n=2$ with $U=8t$ (intermediate $U$
ferrimagnetic state) and $10t$ (large $U$ ferromagnetic orbital ordered
state). Schematic illustrations of sub-band energy structure at $U=8t$ (b)
and $U=10t$ (c). Shaded areas show the continuum state. Broken lines are the
Fermi energy. Remark: band curvature not shown.}
\label{fig:spectra_n2}
\end{figure}

To summarize: the electronic spectrum consists of a number of different
sub-bands (with the detailed structure depending on number of La layers and
on the nature of the ordering, if any). The optical spectrum for $z$%
-polarized light is predicted to consist of relatively sharp peaks, which
may be assigned to the different optically allowed intersubband transitions,
along with broad features (often rather weak) associated with transitions
from sub-bands to the continuum.

\section{Summary and Conclusion}

In this paper we studied theoretically the phase diagram and electronic
properties of a ``correlated heterostructure'' model involving $n$ layers of
a material which in bulk is a Mott insulator, embedded in an infinite band
insulator. The specific features of the model we study were chosen to
reproduce the LaTiO$_3$/SrTiO$_3$ superlattice system studied by Ohtomo 
\textit{et al.}, but we hope that our results will shed light also on the
more general question of the physics of the interface between a strongly
correlated and weakly correlated systems.

A crucial feature of the experimental LaTiO$_3$/SrTiO$_3$ system studied by
Ohtomo \textit{et al.} is the almost perfect lattice match between the two
systems. These authors argued that this implies that the only difference
between the Mott insulating and band insulating regions arises from the
different charge of the La(3+) and Sr(2+); in particular the crystal
structure and atomic positions are expected to remain relatively constant
throughout the heterostructure. Of course, the asymmetry present at the LaTiO%
$_3$/SrTiO$_3$ interface must induce some changes in atomic positions: a TiO$%
_6$ octahedron is negatively charged, and so if it sits between a Sr plane
and a La plane it will be attracted to the latter, and also distorted,
because the positively charged Ti will tend to move in the opposite
direction. The experimentally determined Ti-Ti distances shown in Fig.~1 of
Ref.~\onlinecite{Ohtomo02}, along with the distortion in that paper,
suggests that the changes in Ti-Ti distance are negligible. In this
circumstance, changes in O position along the Ti-O-Ti bond change hoppings
only in second order. We therefore neglected these effects and assumed that
the electronic hoppings and interaction parameters remain invariant across
the heterostructure. However, we emphasize that properly accounting for the
effect of atomic rearrangements inevitably present at surface and interface
is crucial. We further note that lattice distortions appear to be important
in stabilizing the observed bulk state, but may be pinned in a
heterostructure. Extending our results to include these effects is an
important open problem.

In the calculation reported here the heterostructure is defined only by the
difference (+3 vs +2) of the La and Sr charge. The calculated electronic
charge density is found to be controlled mainly by electrostatic effects
(ionic potentials screened by the electronic charge distribution). Results,
such as shown in Fig.~\ref{fig:n_tot} for $U=6t$, are representative of
results obtained for a wide range of on-site interaction $U$. We find
generally that significant leakage of charge into the band insulator region
occurs. The width of the transition region must depend on the relative
strength of the $z$-direction hoppings and the confining potential. For the
parameters studied here, the transition region is about 3 layers (so one
needs about 7 La layers to obtain a central region with bulk behavior), and
that the $n=1$ case is special because the La counter-ions sit between the
Ti planes. We note, however, that the moderate to large $U$-values which we
study do ensure that even for thick superlattices the charge density on the
central layers never becomes greater than unity.

The calculated width is found to be somewhat less than that found by Ohtomo 
\textit{et al.}, broadening our distribution by the experimental uncertainty
(obtained from the La positions) leads to calculated width about 2/3 of
measured ones. Whether the difference arises because we have overestimated
the confining potential or for some other reason such as La/Sr
interdiffusion remains to be determined.

We now turn to our calculated phase diagram shown in Fig.~\ref{fig:diagram}.
It is expected on general grounds that decreasing the number of La layers
will raise the interaction values required for obtaining ordered states.
Further, the specific structure of the system of present interest implies an
odd-even alternation. Both of these features are indeed observed in our
calculation. Our calculation, in combination with the $U$-values inferred
from optics, suggests that the La/SrTiO$_3$ system may be in the interesting
($U \sim 7t \mbox{-} 12t$) parameter regime in which one or more phase
boundaries may be crossed by varying La layer number. A further point,
perhaps especially relevant to systems such as titanates which allow for
orbital ordering and orbitally dependent hopping, is that the structural
anisotropy intrinsic to the heterostructure may favor different ordering
patterns than those found in bulk. Thus we find for the thinnest
heterostructures different phases with more translational invariance than
found in bulk. We note, however, that in the Ti case the ``edge states''
have a quasi one-dimensional character with an incommensurate filling,
perhaps favoring incommensurate charge, spin or orbital ordering at the edge.

The issue of the transport properties of the heterostructure is an important
open question, especially in light of the interesting transport data of Ref.~%
\onlinecite{Ohtomo02}. A crucial experimental finding is that metallic
conduction is always observed, consistent with our prediction of conducting
edge states. However, we do not have a qualitative understanding of the Hall
data, because within the Hartee-Fock approximation the Hall resistivity is
obtained by adding up the contributions of the different sub-bands. The
contributions to $\sigma_{xy}$ arising from the quasi one-dimensional $xz,yz$
sub-bands are controlled by weak deviations from the nearest neighbor
hopping approximations, which control the mixing between the two bands and
the reconstruction of the Fermi surface of the crossing point. It is
difficult to make general statements without more detailed band structure
information, but some degree of compensation is expected.

Other important issues for future research include study of correlation
effects beyond the Hartree-Fock approximation, inclusion of the coupling
between atomic rearrangements, orbital ordering and electronic hopping
parameters, and more sophisticated treatment of the dielectric properties of
SrTiO$_3$.

\textit{Acknowledgements} We acknowledge very helpful conversations with H.
Hwang and M. Potthoff. This research was supported by NSF DMR-0338376
(A.J.M.) and JSPS (S.O.).


\begin{thebibliography}{99}
\bibitem{U_dependence} Actually, the crossover length is dependent on both
the parameters $\varepsilon$ and $U$ in the following way; with increasing $%
\varepsilon$, the confinement potential due to the La layer becomes weak and
the crossover length becomes longer, with increasing $U$ the electron
density at each site decreases thus the crossover length becomes longer.
However, $\varepsilon$ and $U$ dependences are weak, 
and qualitative behavior does not change.
\end{thebibliography}
\end{document}